\documentclass[a4paper,twocolumn,superscriptaddress,nogroupedaddress,accepted=2023-03-15]{quantumarticle}
\pdfoutput=1

\usepackage[numbers,sort&compress]{natbib}
\usepackage{xcolor,graphicx}
\usepackage{amssymb,amsmath,graphicx}
\usepackage[utf8]{inputenc}
\usepackage{comment}
\usepackage{braket}
\usepackage[normalem]{ulem}
\usepackage{hyperref}
\usepackage{url}
\usepackage{array}
\usepackage{fontawesome}

\begin{document}

\title{Cooperative quantum information erasure}

\author{Lorenzo Buffoni}
\address{Portuguese Quantum Institute, Lisbon, Portugal}
\address{Department of Physics and Astronomy, University of Florence, 50019 Sesto Fiorentino, Italy}
\author{Michele Campisi}
\address{NEST, Istituto Nanoscienze-CNR and Scuola Normale Superiore, I-56127 Pisa, Italy}

\begin{abstract}
We demonstrate an information erasure protocol that resets $N$ qubits at once. The method displays exceptional performances in terms of energy cost (it operates nearly at Landauer energy cost $kT \ln 2$), time duration  ($\sim \mu s$) and erasure success rate ($\sim 99,9\%$).
The method departs from the standard algorithmic cooling paradigm by  exploiting cooperative effects associated to the mechanism of spontaneous symmetry breaking which are amplified by quantum tunnelling phenomena. Such cooperative quantum erasure protocol is experimentally demonstrated on a commercial quantum annealer and could be readily applied in next generation hybrid gate-based/quantum-annealing quantum computers, for fast, effective, and energy efficient initialisation of quantum processing units.
\end{abstract}

\maketitle

\section{Introduction}
Preparing large sets of qubits in a pre-specified pure state is of crucial importance for the achievement of powerful quantum computers \cite{Preskill18QUANTUM2,Cirac99PRL82,buffoni2022third,fellous2021limitations}. This has led to the proposal and realisation of a number of ``algorithmic cooling'' techniques whereby an entangling unitary operating on many qubits being initially in a mixed state results in the purification of a subset thereof  \cite{Schulman99STOC99,Fernandez04IJQI02,Schulman05PRL94,Baugh05NAT438,Solfanelli22AVSQS4}. In view of the rising concerns regarding the energetic footprint associated to massive computation that characterises our modern society, and the possibility that quantum computers could alleviate it \cite{Auffeves22PRXQ3}, it is currently becoming clear that is of crucial importance that protocols of multiple qubit reset be devised and practically demonstrated that not only are effective, but also fast and energy efficient. 

According to Landauer's principle \cite{Landauer61IBMRD5}, resetting a single bit from a random state to a prespecified state (the so-called erasure of one bit of information) costs at least $kT \ln 2$ of work, where $T$ is the temperature of the environment that surrounds the register. 
It has been established in recent years that the same bound holds as well for resetting a single qubit from completely mixed state to a prespecified pure state \cite{Piechocinska00PRA61,Esposito10NJP12}. 
As demonstrated in a number of experiments with classical registers \cite{Berut12Nature483,Berut13EPL103,Berut15JSTAT,Gavrilov16PRL117,Gavrilov17PNAS114,Proesmans20PRL125,Saira20PRR2,Buffoni22JSP186} the Landauer bound is typically achieved at the cost of longer and longer duration, so a relevant figure of merit that quantifies the performance of an information erasing device is the product of its energy cost per bit $w$ and its time duration $\mathcal{T}$ \cite{Likharev77IEEETM13,Gaudenzi18NATPHYS14}
\begin{align}
\mathcal A = w \mathcal{T}
\end{align}
for which we coin the expression ``erasure action''. The smaller $A$ the better the device.
In an earlier work, Gaudenzi \emph{et al.} \cite{Gaudenzi18NATPHYS14} have demonstrated that quantum tunnelling can enable fast and maximally efficient reset which may result in the record erasure action $\mathcal A$ of the order of $10^{-22}$ erg$\cdot$s/bit using molecular nano-magnets. 

Here we experimentally demonstrate a simple and effective method to erase $N$ qubits at once, whose erasure action $\mathcal A$ is on the order of  $10^{-23}$ erg$\cdot$s/bit while operating (within the estimated error) at the Landauer limit, and with an erasure success rate of $99,9 \%$.

What makes all this possible is a shift in the cooling paradigm: here the peculiar physics of spontaneous symmetry breaking (SSB) is employed to generate a cooperative many body effect among the $N$ qubits that is amplified by the enhancing action of quantum phenomena. 

Our experimental demonstration uses a remotely programmable NISQ device \cite{Preskill18QUANTUM2}, specifically a quantum annealer, as the experimental platform. Accordingly, the results presented should not be considered as simulations of physical phenomena, or as theoretical results. As discussed below, $N$ physical qubits were actually cooled from a completely mixed state, to a nearly pure state in our experiments. The degree of remote control on the hardware permitted not only to realise that, but also to estimate the erasure energy cost, time duration, and rate of success.

\section{The main idea}

\begin{figure}[t]
    \centering
    \includegraphics[width=0.6\linewidth]{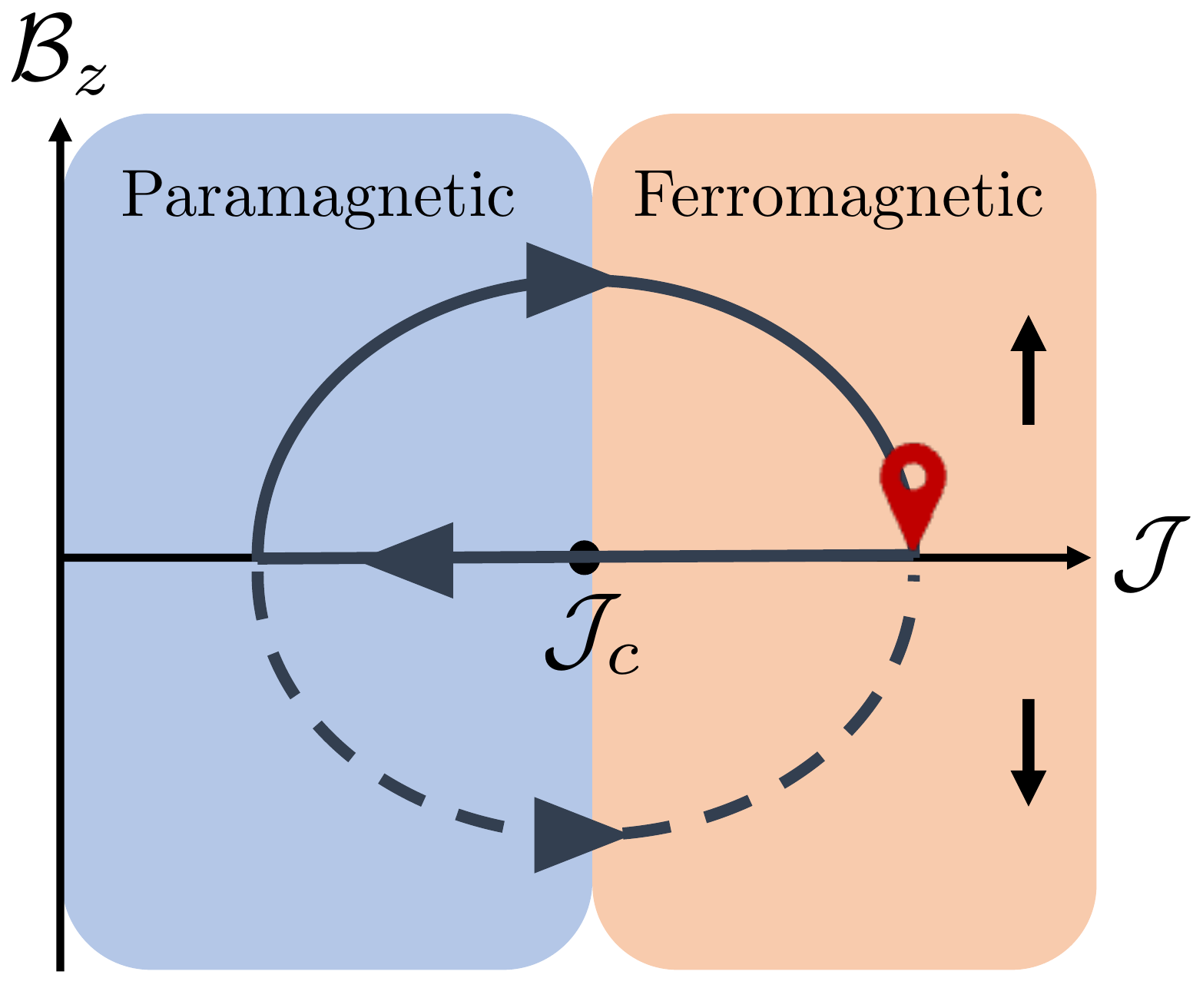}
    \caption{Landauer erasure protocol based on spontaneous symmetry breaking (classical version). We implemented it, and as well a quantum enhanced version thereof, on a commercial quantum annealer.}
    \label{fig:parrondo}
\end{figure}

\begin{figure*}[t]
    \centering
    \includegraphics[width=\textwidth]{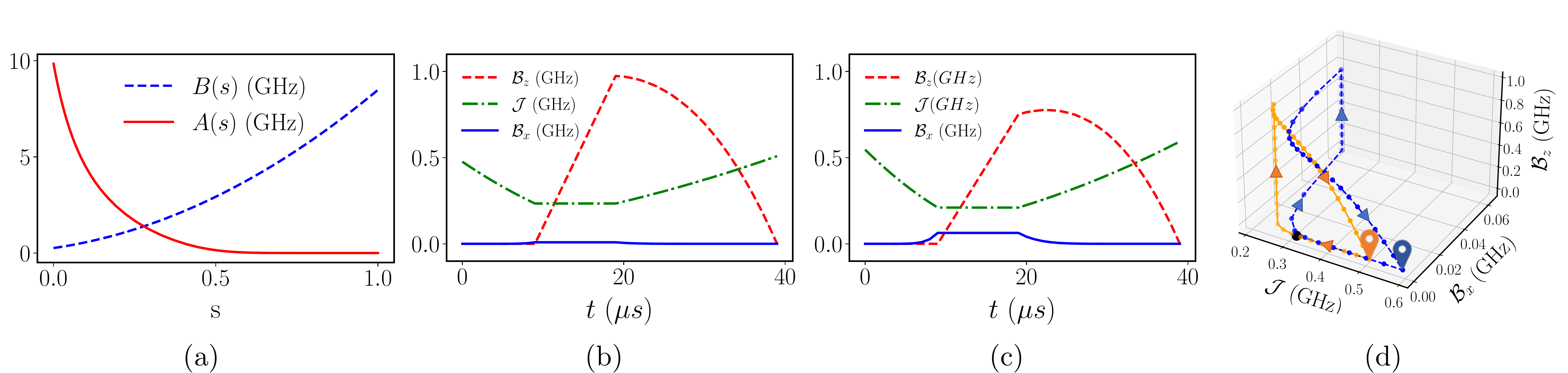}
    \caption{(a) Control fields $A(s)$ and $B(s)$ of the D-Wave Hamiltonian, Eq.( \ref{eq:advantage_ham}), as a function of the annealing parameter $s$. (b) Driving fields for the classical erasure protocol as a function of time. (c) Driving fields for the quantum erasure protocol as a function of time. (d) Classical (orange solid line) and quantum (blue dashed line) erasure protocols depicted as paths in the space of driving fields. The dots along the paths mark the times $k\delta t$, with $k=0\dots 40$, $\delta t=1\mu$s. The black dot roughly indicates the position of the critical point and pins mark the start and end point of the erasure schedules.}
    \label{fig:protocols}
\end{figure*}

Our information erasure strategy is based on the protocol illustrated in Fig. \ref{fig:parrondo} .
It is inspired by the Szilard engine protocol of Ref.  \cite{Parrondo01CHAOS11}, which traverses the same path, but in the opposite  direction. 
The register is a uniaxial Ising ferromagnet immersed into an environment at some temperature $T$  \cite{GoldenfeldBook}. Imagine the spin-spin interaction energy $\mathcal J$ is such that the register is in the ferromagnetic state and there is no external applied magnetic field, $\mathcal B_z$. Due to the mechanism of SSB the register has a net magnetisation either pointing up (logical state \texttt{1}) or down (logical state \texttt{0}) with same probability.
Now if you decrease the interaction energy $\mathcal J$, when you cross the  critical value $\mathcal{J}_C$ the system gets into the paramagnetic state and becomes very sensitive to an external field. By circumventing the critical point from above (below) in the 
$( \mathcal J,\mathcal B_z)$ plane, and return to the ferromagnetic state, the magnet will result in a positive (negative) net magnetisation, regardless of its initial magnetisation. That is, thanks to the peculiar physics of SSB one can map both logical states $0$ and $1$ onto the single logical state $1$ (or $0$ depending on the chosen path), which is Landauer's erasure of one bit of information. 

It is important to note that the above described erasure protocol was never experimentally implemented so far. One difficulty is evidently that manipulating the interaction energy $\mathcal J$ of a physical magnet is not practically feasible. Using a quantum annealer as our experimental platform allowed to overcome that difficulty and achieve, for the first time, the implementation of the SSB based protocol of Fig. \ref{fig:parrondo}.

Notwithstanding the interest of such implementation on its own, that does not constitute the main result reported in this work, but only the basis for the discovery of further new phenomena, which were never observed nor predicted before.
Besides the SSB based protocol of Fig. \ref{fig:parrondo} we also implemented a quantum version thereof, featuring the turning on of quantum tunnelling between the two logical states which resulted in a faster and more energy efficient erasure.
Furthermore, the quantum version of the protocol was observed to be so effective that it actually erases the information carried by each single spin -- a new phenomenon which we dub ``cooperative quantum information erasure'' of $N$ qubits at once. Its discovery, experimental demonstration and observed exceptional performance (in terms of time, energy consumption and success rate) is the main result of the work.

\section{Implementation}

We experimentally implemented our SSB based information erasure protocols on the  D-Wave Advantage 4.1  processor \cite{dwdocs}. It consists of a network of coupled superconducting qubits, whose dynamics is well described by an Ising Hamiltonian in transverse field
\begin{equation}
    \frac{H(s)}{h} = -\frac{A(s)}{2}\sum_i \sigma^x_i - \frac{B(s)}{2}\left [ g\sum_i \sigma^z_i +  \sum_{i,j } J_{ij} \sigma^z_i\sigma^z_j \right ]
    \label{eq:advantage_ham}.
\end{equation}
Here $h$ is Planck's constant, $\sigma^\alpha_i$ denote Pauli operators, and the parameters $A$ and $B$, with dimension of frequency, are predetermined functions of the so called annealing parameter $s$, see Fig. \ref{fig:protocols}(a). The annealing parameter $s$ can be manipulated in time by the user according to some piece-wise linear function of time $s(t)$. This makes the Hamiltonian generally a time-dependent operator. Users have as well some degree of control of the parameter $g$ in time. There is also some freedom in choosing the $J_{ij} $'s but their value is fixed in time.

In our implementation, the network geometry was that of a $16 \times 16$ square lattice, with constant nearest neighbour interaction $J_{ij}=J$, thus resulting in a Hamiltonian of the form 
\begin{equation}
  \frac{H(t)}{h} =  -\mathcal{B}_x(t)\sum_i \sigma^x_i - \mathcal{B}_z(t)\sum_i\sigma^z_i - \mathcal{J}(t)\sum_{\langle i,j\rangle} \sigma^z_i\sigma^z_j
    \label{eq:ham}
\end{equation}
where $ \mathcal{B}_x(t)= A(s(t))/2$, $\mathcal{B}_z(t)= B(s(t))g(t)/2$, $\mathcal J(t) = B(s(t))J/2$. The choice of bi-dimensional geometry, and relatively large $N$ was dictated by the necessity of the system to display  a classical SSB (there is no SSB in the classical 1D Ising model) \cite{GoldenfeldBook}.
Clearly, the fact that the parameters $ \mathcal{B}_x, \mathcal{B}_z, \mathcal J $ depend on time through predetermined functions of the annealing parameter, implies that a user does not have full freedom to control them, independently. 
However, given the hardware constraints, there is enough room to implement a purely classical information erasure protocol (i.e., one with $ \mathcal{B}_x(t) \simeq 0$, where all terms in the Hamiltonian commute with each other) of the type in Fig. \ref{fig:parrondo}, and, as well a quantum version thereof featuring non null transverse field $ \mathcal{B}_x(t)$ that enables quantum tunnelling between the two wells of the Landau free-energy potential \cite{GoldenfeldBook}.
 
Figures \ref{fig:protocols}(b) and \ref{fig:protocols}(c) respectively show the classical  and quantum erasure schedules $ \mathcal{B}_x(t), \mathcal{B}_z(t), \mathcal J(t)$, used in our experiments, which resulted from a careful choice  of the functions $s(t)$ and $g(t)$ and the parameter $J$.
Figure \ref{fig:protocols}(d) shows the classical (orange) and quantum (blue) erasure protocols as closed paths in the ($ \mathcal{B}_x, \mathcal{B}_z, \mathcal J $) space. Note how the classical protocol 
implements an information erasure scheme of the type depicted in Fig. \ref{fig:parrondo}, and that in the quantum protocol a $ \mathcal{B}_x$ field is turned on around the crossing of the critical point $ \mathcal{B}_x=0, \mathcal{B}_z= 0, \mathcal J = \mathcal J_C$, which we roughly estimate to be $\mathcal J_C \sim 0.36$ GHz.

\begin{figure}[ht]
    \centering
    \includegraphics[width=0.7\linewidth]{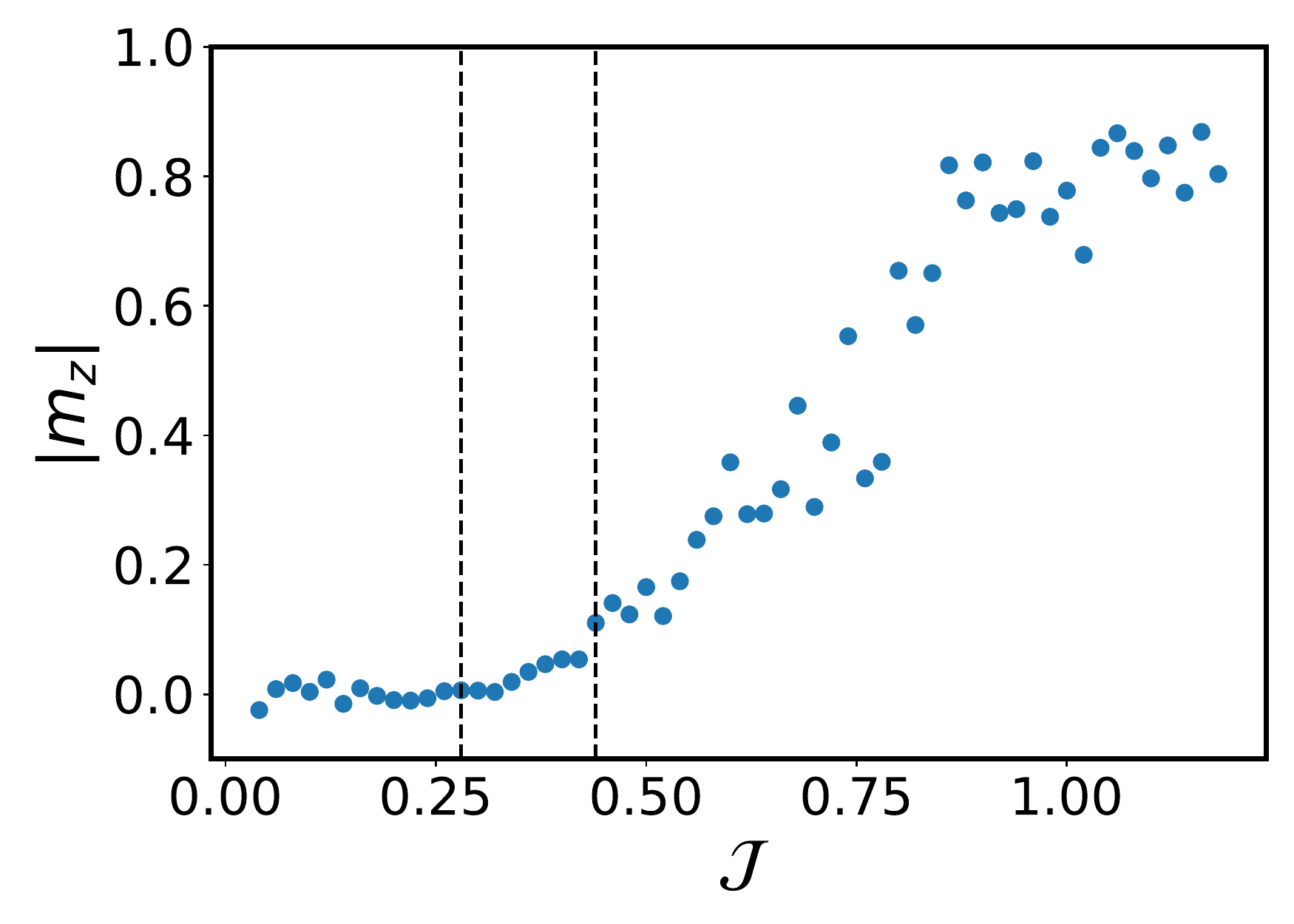}
    \caption{Modulus of magnetization density $|m_z|$ as a function of the coupling $\mathcal{J}$. Each point is the average over 1000 samples. Since the system presented a weak non-zero magnetization even at zero coupling we have offset that value to generate this plot. The vertical lines denote the region around which the spontaneous magnetization starts to grow and thus the phase transition is located. We pick the middle of this interval as our $\mathcal{J}_C$ value and its half-width of the interval as its uncertainty.}
    \label{fig:jc}
\end{figure}

To perform the estimate of $\mathcal J_C$ we run several slow forward annealing ramps $s(t)=t/\tau$ at $g=0$ and $J_{ij}=J$, thus evolving the Hamiltonian from $ H(0)/h = - A(0)\sum_i \sigma^x_i/2 $ to $ H(1)/h =  -  B(1) J \left [ \sum_{i,j } \sigma^z_i\sigma^z_j \right ]/2
$, in a time $\tau$. We then measured the final magnetisation density (see Eq.(\ref{eq:magnetisation-density}) below for its definition) which correspond to the equilibrium distribution of the Ising model.  We took $\tau=200 \mu s$, which was slow enough to sample from the equilibrium distribution.\footnote{Annealing runs featuring larger $\tau$ yielded same results.} We repeated this process for different values of the coupling $J$. Fig. \ref{fig:jc} shows the so measured absolute value of the magnetisation density as a function of $\mathcal J=B(1)J/2$. The critical point was then estimated as the point above which we observed a noticeable deviation from zero of the magnetization curve  (see caption of Fig. \ref{fig:jc} for more details). 

The estimated $\mathcal{J}_C\sim 0.36 \pm 0.08$ GHz corresponds to an estimated temperature of the qubit system $T=39\pm 9$ mK, according to the formula $kT = 2 \mathcal J_C/\ln(1+\sqrt{2})$ \cite{GoldenfeldBook}, which is about twice the D-Wave cryostat nominal temperature of $15.4$ mK.\footnote{As a double check, we also tried to estimate the temperature with the pseudo-likelihood method detailed in \cite{benedetti2016estimation}.
This method turned out to be not very reliable as it has the drawback of providing estimates that may vary considerably as  $\mathcal J$ is varied (especially when comparing estimates performed in the paramagnetic vs. the ferromagnetic phase). This notwithstanding, all so performed estimates taken in the chosen range of Fig. \ref{fig:jc}, were consistent with the interval of $T=39\pm 9$ mK, which thus remains our best estimation of temperature.}

\section{Erasure of a single macroscopic bit}

\begin{figure*}[ht]
    \centering
    \includegraphics[width=0.8\linewidth]{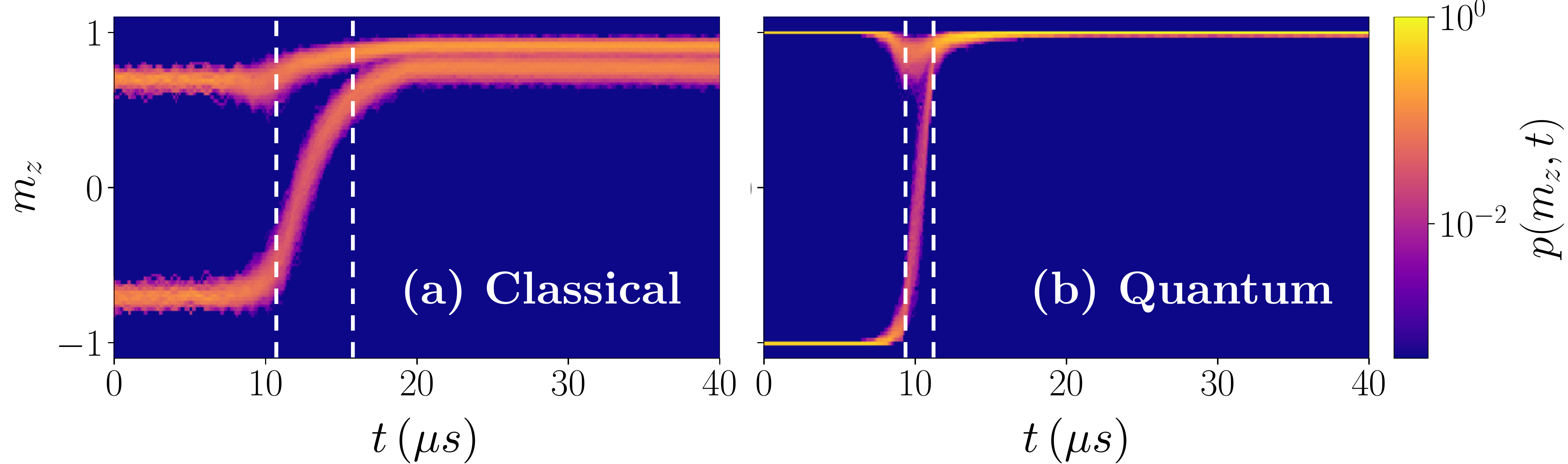}
    \caption{Probability $p(m_z,t)$ to observe a magnetisation density $m_z$ at time $t$ for the classical (a) and quantum (b) erasure of a single macroscopic bit, as obtained from the quantum annealing experiments. The vertical dashed lines mark the beginning and ending of the erasure process.}
    \label{fig:traj}
\end{figure*}

Figure \ref{fig:traj} shows the results regarding the erasure of a single macroscopic bit, according to the main idea described in Fig. \ref{fig:parrondo}, and its quantum version. In those experiments the whole Ising system is seen as a single memory register that holds one bit of information. Its logical state is $\texttt 0$ ($\texttt 1$) if it has a net magnetisation density 
\begin{align}
m_z = {\sum_i s_i}/{N}
\label{eq:magnetisation-density}
\end{align}
below (above) some threshold value $-\Delta$ ($+\Delta$). The symbols $s_i$ denote the eigenvalues of $\sigma^z_i$. The plots depict the probability $p(m_z,t)$ that the system has a magnetisation density $m_z$ at time $t$, as obtained from the quantum annealing output data, for the classical erasure protocol (panel a), and the quantum protocol (panel b). 

To generate the plots in Fig. \ref{fig:traj} we first initialise the D-Wave Advantage 4.1 \cite{dwdocs} quantum annealing processor in a classical configuration $\{ s_i \}$, meaning that for $i=1 \dots N$, qubit $i$ was prepared in the eigenstate $|s_i\rangle $ of $\sigma_z^i$ corresponding to the eigenvalue $s_i$. 
The configuration  $\{ s_i \}$ is randomly sampled from an initial distribution $P_0(\{s_i\})$\footnote{Accordingly, the density operator that describes the initial state of each run is $\rho(\{ s_i\})= \otimes_i  |s_i\rangle \langle s_i|$, and that occurs with a probability $P_0(\{s_i\})$. Namely, the density operator that describes the initial state of the whole statistical ensemble of our experimental runs is the mixture $\rho= \sum_{\{ s_i\}} P_0(\{s_i\} \rho(\{ s_i\})$.}.

Note that we used two distinct initial distributions $P_0(\{s_i\})$ for the quantum and classical cases (Fig. \ref{fig:traj}, panel a) and panel b), respectively). We provide details on the chosen $P_0(\{s_i\})$'s below.

Once initialised in a given configuration $\{ s_i \}$, we let the system evolve under the action of the chosen erasure schedule (either the classical or the quantum one) for a time $t$, and measure the observable $\sigma_i^z$ for each of the $N =16 \times 16=256$ qubits.
That gives the evolved  configuration $\{ s_i^t \}$ and the according magnetisation density, Eq. (\ref{eq:magnetisation-density}), at time $t$.
This procedure was repeated $\mathcal N = 2000$ times for each erasure protocol in order to collect configuration statistics.
The frequency that a certain value $m_z$, Eq. (\ref{eq:magnetisation-density}), is observed at time $t$, gives an estimate of the probability $p(m_z,t)$ to find the total system with a magnetisation density $m_z$ at time $t$. This was done for $t= k \delta$, with $\delta = 1\mu s$, and $k=1, \dots, 40$, to provide a picture of how the probability $p(m_z,t)$ evolves in time for the two protocols, i.e., the plots in Fig. \ref{fig:traj}. 

Let us now turn to illustrate our choice of sampling distributions $P_0(\{s_i\})$. 
For the classical cases (Fig. \ref{fig:traj}, panel a) we defined  $P_0(\{s_i\})$ in the following manner. We prepared the annealer in the configuration where all qubits are aligned and pointing in the negative direction $s_i= - 1$. We let this configuration evolve for a time $t$ according to the classical erasure protocol, and repeat $\mathcal N = 2000$ to collect statistics, for various times $t$ up to $40\mu s$, as described above. We observed the crucial fact that the configuration distribution converges to a steady state distribution $P_+(\{s_i\})$, associated with a magnetisation density distribution $p_+(m_z)$ featuring a single narrow peak located well within the region that well encodes the logical state 1. Accordingly the mirror image $P_-(\{s_i\})=P_+(\{-s_i\})$
well encodes the logical  state 0. The initial distribution of the classical plot in Fig. \ref{fig:traj} is thus sampled from the even combination of logical 0 and logical 1, i.e., $P_0(\{s_i\})=[P_+(\{s_i\})+P_-(\{s_i\})]/2$. This is reflected in the initial magnetisation density distribution being doubly peaked.

For the quantum case  (Fig. \ref{fig:traj}, panel b) we proceeded in the same way as above, only now the quantum erasure protocol was employed. We observed that now the steady state $P_+(\{s_i\})$ is very close to the distribution, call it
$\delta_+(\{ s_i\})$, featuring all spins up. We thus now chose $P_0(\{s_i\})$ as the even combination of $\delta_+(\{ s_i\})$ and $\delta_-(\{ s_i\})=\delta_+(\{ -s_i\})$: $P_0(\{s_i\})=[\delta_+(\{ s_i\})+\delta_-(\{s_i\})]/2$. This is reflected in the initial magnetisation density distribution being the sum of two delta peaks located at maximal and minimal magnetisation.

The plot in Fig. \ref{fig:traj}(a) evidences the first ever reported successful experimental implementation of the SSB  based information erasure protocol, depicted in Fig. \ref{fig:parrondo}.  The initial distribution featuring two well separated peaks encoding two distinct logical states $\texttt 0$ and $\texttt 1$, ends up in a distribution with two very overlapped peaks that encode one and the same logical state $\texttt 1$.

The  plot in Fig. \ref{fig:traj}(b) evidences further new phenomena, namely the experimental implementation of the quantum version of  SSB  based information erasure. In this case as well the erasure is successful. A striking observations is that such quantum version is faster than the classical one. A detailed quantitative analysis, offered below in Sec. \ref{sec:analysis},  shows as well that it is also more energy efficient. Most remarkably, we observe that the quality of the distribution that encodes the logical states in the quantum case is extremely higher than that of the classical case. While the magnetization density expectation at the end of the erasure protocol is $\langle {m_z} \rangle =0.84 \pm 0.08$ in the classical case, in the quantum case it is  $\langle m_z \rangle =0.998 \pm 0.004$. This means that $99.9 \%$ spins were aligned in the positive $z$ direction at the end of the quantum erasure. This suggests that the quantum method not only efficiently erases the information at the macroscopic level, but it efficiently erases the information even at the fine level of each microscopic constituent, i.e., each qubit. This intuition was corroborated by further experiments described below in Sec. \ref{sec:coop}.

\begin{figure*}
    \centering
    \includegraphics[width=0.8\linewidth]{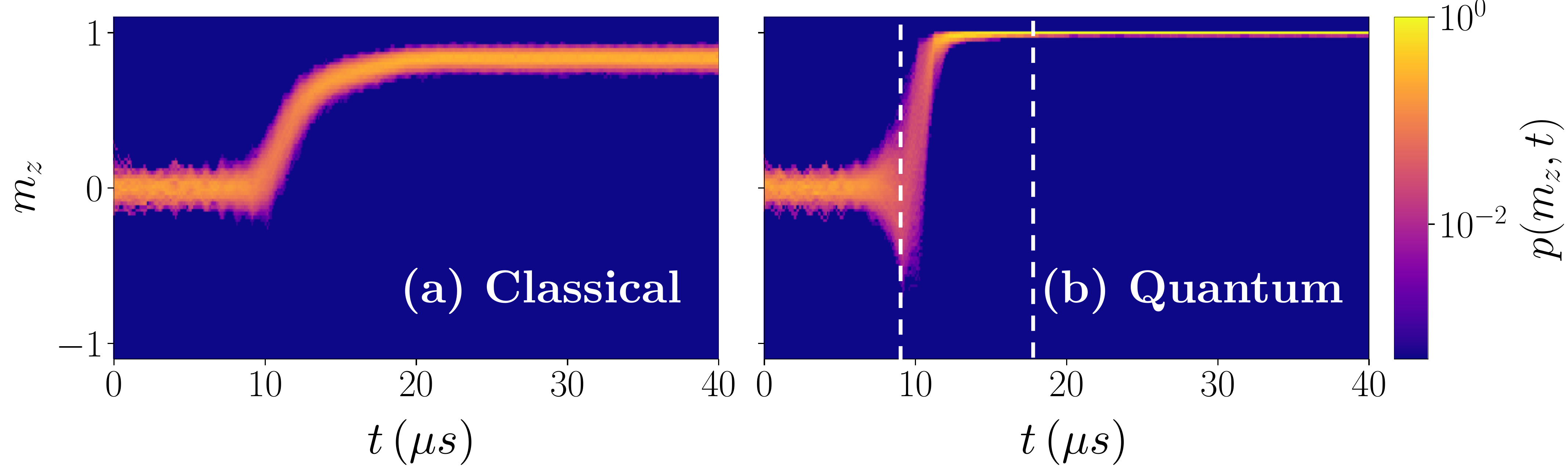}
    \caption{Probability $p(m_z,t)$ to observe a magnetisation density $m_z$ at time $t$ for the classical (a) and quantum (b) erasure of $N$ qubits at once, as obtained from the quantum annealing experiments. The vertical dashed lines mark the beginning and end of the collective erasure process. }
    \label{fig:random}
\end{figure*}

\section{Cooperative quantum information erasure}
\label{sec:coop}
 
To corroborate the intuition that  the quantum protocol is able to erase information at a single qubit level,
we performed a second set of experiments where each qubit of the lattice was initialised in a completely random way, that is the initial configuration distribution $P_0(\{s_i\})$ is now the flat distribution over configuration space $P_0(\{s_i\})=2^{-N}$. This is reflected in the initial magnetisation density distribution $p(m_z,0)$ being now bell shaped with zero mean. We applied the flat initialisation to both the classical and quantum erasure schedules.

At variance with the previous experiment, in this second one the whole system is seen as a set of $N$ independent microscopic registers, each carrying one bit of information, and not as one macroscopic bit that carries information in a collective degree of freedom. The erasure protocols, however are just the same used above and depicted in Fig \ref{fig:protocols}; what changes is only the initial preparation (that reflects the fact that the qubits are now seen as independent registers) and the interpretation of results. 

Figure \ref{fig:random} reports the time evolution of the magnetisation density $p(m_z,t)$, which was computed from the output data in the same way as described for the first experiment. Note how in the quantum case, almost every single qubit is taken from its initial random state to the up state, which corroborates the intuition that information is indeed erased at the fine level of each qubit. Indeed the quantum erasure has a final average magnetization of $\langle m_z \rangle =0.999 \pm 0.003$ and the classical one of $\langle m_z \rangle =0.84 \pm 0.03$.

What we are observing is rather surprising: the quantum phenomena greatly amplify the cooperative SSB phenomena thus resulting in the  erasure of many qubits at the same time. We are witnessing a cooperative quantum information erasure phenomenon.

It is important to remark that because of SSB, the end state is very stable in the course of time. To further corroborate this fact, we performed further runs with the final distribution $p(m_z,\tau)$ of the quantum protocol as the initial state of a new evolution  lasting $2000 \mu$s (the maximum allowed by the hardware) with the fixed Hamiltonian $H(0)$: no change in the distribution was observed over the course of that time.

\section{Data analysis}
\label{sec:analysis}

\subsection{Erasure time}

To assess the speed of the two protocols in Fig. \ref{fig:traj} we proceeded in the following way. We first introduce a threshold $\Delta$ that defines the two logical states of our ``magnet'', meaning that a magnetisation density $m_z$ above $\Delta$ denotes the logical $1$, and a value below $-\Delta$ denotes the logical $0$. We chose (with a good degree of arbitrariness) the value $\Delta =0.5$.  Now, considering the switching events only (namely those that began in the logical state $0$, i.e., the lower branches in the plots of Fig. \ref{fig:traj}), we define the switching time as the time that it takes for the average magnetisation $m_z$ to switch from the logical state $0$, to the logical state $1$. The vertical dashed lines in the Fig. \ref{fig:traj} denote the times at which the average $m_z$ crosses the values $\pm \Delta$, and their temporal distance is our measure of the switching time. Despite our  limited temporal resolution of $1 \mu s$, it is clear that the quantum erasure process is faster than the classical one. The switching time of the quantum protocol is around $2\mu s$ in contrast with the $5 \mu s$ required by the classical one.

To assess the speed of the cooperative quantum erasure depicted in Fig. \ref{fig:random}b), we proceeded in a similar way as with Fig. \ref{fig:traj}. We now just set the threshold $\Delta$ to the asymptotic value $0.998$ (i.e.,  $99.9\%$ of qubits pointing up), and set the start of the switching time as the instant when the transverse field is turned on. This gave an estimate of switching  time, at $99,9 \%$ success rate, of
\begin{align}
\mathcal T_{99,9 \%} = 9 \mu s
\end{align}
We remark that obviously the estimation of the switching time depends on the arbitrary choice of the magnetisation threshold $\Delta$. For example with $\Delta = 0.98$ (i.e.,  $99\%$ of qubits pointing up) the switching time drops to $3 \mu s$, while the value $\Delta = 0.99$ (i.e., success rate of  $99,5\%$) is reached in about $5 \mu s$. The work value would drop accordingly if one is happy with a lower success rate. This evidences that there is a trade-off between erasing action and success rate.

\subsection{Erasure work}

Given a path $\mathcal C$ in the space $ \mathcal{B}_x, \mathcal{B}_z, \mathcal J$ and the according values taken 
by their respective conjugated ``forces'', namely 
\begin{align}
M_x=\sum_i \langle \sigma^x_i \rangle,\, M_z= \sum_i \langle \sigma^z_i \rangle,\, K= \sum_{\langle i,j \rangle} \langle  \sigma^z_i \sigma^z_j \rangle
\end{align}
where the symbol $\langle \cdot \rangle$ denotes quantum mechanical expectation,
the average work cost of the process $\mathcal C$ reads
\begin{align}
W &= - h \int_{\mathcal C} \left(M_x  d \mathcal{B}_x + M_z  d\mathcal{B}_z + K d\mathcal{J}\right) \\
&\doteq W_x + W_z +W_{zz}. \nonumber
\label{eq:totalW}\,
\end{align}

For all four experiments depicted in Figs. \ref{fig:traj}, \ref{fig:random}, at each point along the path we have collected the eigenvalue of each $\sigma_i^z$ operator, and repeated the measurement $\mathcal N$ times. By averaging over such repetitions gave us an estimate of the expectation values $\langle \sigma^z_i \rangle$ and $\langle  \sigma^z_i \sigma^z_j \rangle$. Hence we have a sampling of the forces $M_z$ and $K$ along our paths which could be used to estimate the according works $W_z$ and $W_{zz}$ as discrete sums.

Regarding the term $W_x$, its estimation is problematic because the hardware does not allow to measure the magnetisation along the x-direction. However we could estimate a quantity $\delta W$
that bounds its modulus. Recalling that the quantities $\langle \sigma^x_i \rangle$, $\langle \sigma^y_i \rangle$, $\langle \sigma^z_i \rangle$, are the components of the Bloch vector describing the state of the qubit $i$, and that said vector has at most length $1$, we have
\begin{align}
| M_x | =& | \sum_i \langle \sigma^x_i \rangle |  \leq 
 \sum_i |\langle \sigma^x_i \rangle |
\leq 
\sum_i \sqrt{1-\langle \sigma^z_i \rangle^2} \doteq M_* \nonumber
\end{align}
where the last term is accessible from the quantum processor output data.
Let's break $W_x$ in two contributions, one stemming from the forward branch $ \mathcal{B}_x^\text{min}\rightarrow   \mathcal{B}_x^\text{max}$, and one from the backward branch $  \mathcal{B}_x^\text{max}\rightarrow  \mathcal{B}_x^\text{min}$ and let $ M_x ^{F(B)}, M_*^{F(B)}$ denote the according average magnetisations and their bounds. We have:
 \begin{align}
&W_x =- h\int_{ \mathcal{B}_x^\text{min}}^{  \mathcal{B}_x^\text{max}} M_x ^F d  \mathcal{B}_x -  
h\int_{  \mathcal{B}_x^\text{max}}^{ \mathcal{B}_x^\text{min}} M_x ^B d  \mathcal{B}_x \nonumber \\
&= - h\int_{ \mathcal{B}_x^\text{min}}^{  \mathcal{B}_x^\text{max}}  (M_x^F  - M_x ^B) d  \mathcal{B}_x\\
&\leq h \int_{ \mathcal{B}_x^\text{min}}^{  \mathcal{B}_x^\text{max}} (| M_x ^F|+ | M_x ^B|) d  \mathcal{B}_x \nonumber \\
&\leq  h\int_{ \mathcal{B}_x^\text{min}}^{  \mathcal{B}_x^\text{max}} (M_*^F + M_*^B) d  \mathcal{B}_x \doteq \delta W\, . \nonumber
\end{align}
Similarly, $W_x \geq -\delta W$, thus
$
- \delta W \leq W_x \leq \delta W
$.
Accordingly our estimation of the total work reads
\begin{align}
W_\text{exp} = W_z + W_{zz} \pm \delta W
\label{eq:Wexp}
\end{align}
where $\delta W$, which is accessible from the experiment data, plays the role of an error caused by the inability to measure $W_x$. 

Equation (\ref{eq:Wexp}) gives directly the energy cost of the single bit erasures depicted in Fig. \ref{fig:traj}(a,b).
Table \ref{tab:tab1} (columns 2 and 3) reports the according measured values of $W_z, W_{zz}, \delta W$ and $W_\text{exp}$.
The most important point to be noted is that the quantum process costs roughly half the work spent by the classical process. Also note that both processes cost far more than the Landauer cost of one bit erasure $W_L= kT \ln 2$.

\begin{table*}
\centering
\begin{tabular}{ | c ||r | r || r |} 
  \hline
  erg $\times 10^{-18}$ & Classical er.  &Quantum er. & Qu. coop. er.\\ 
  \hline   \hline
  $W_z$  & 1067(80) & 331(66) & 166(60)   \\ 
  \hline
  $W_{zz}$ & -351(86) & -9.3(13) & -1140(20)\\ 
  \hline
  $\delta W$  & 36 & 53 & 106\\ 
  \hline
  $U_f$ & & & -1884  \\ 
   \hline
  $W_\text{exp}$  & 716(202) & 322(132) & 910(186)\\ 
  \hline
  $W_L$ & 3.73(0.86) & 3.73(0.86) & 956(220)\\ 
  \hline
\end{tabular}
\caption{Work contributions for the classical erasure, the quantum erasure (Fig. \ref{fig:traj}, panels (a) and (b), respectively), and the quantum cooperative erasure (Fig. \ref{fig:random}, panel (b)). The uncertainties are reported in parenthesis.}
\label{tab:tab1}
\end{table*}

Estimating the work cost of quantum cooperative erasure in Fig. \ref{fig:random}(b) in a way that could be meaningfully be compared to Landauer bound 
$kT \ln 2$ requires some extra care. We recall that the Landauer bound follows from the assumption of an initial thermal state and a cyclical protocol \cite{Buffoni22JSP186},
thus only a protocol of such a type can be meaningfully compared to Landauer erasure and its energy cost be compared to Landauer's cost, accordingly. 
Our initialisation (each qubit is either up or down with same probability) corresponds to the thermal equilibrium of $N$ non interacting qubits in absence of local fields, which are described by the null Hamiltonian $H_0=0$. Thus we consider $H_0$ as the Hamiltonian describing the system, before and after the time dependent protocol $H(t), t \in (0,\tau)$. This means that we have to consider an instantaneous quench from $H_0$ to $H(0)= -h \mathcal J(0) \sum \langle \sigma_i^z \sigma_j^z \rangle$ at $t=0$, and a quench back from $H(\tau)=H(0)$ to $H_0$ at $t=\tau$. The cost of these quenches must be added to the work balance, in order to make a proper comparison with Landauer minimal cost.

For the first quench, which is happening at the beginning of the protocol when the interaction Hamiltonian is turned from $H_0=0$ to $H(0)$, the energy expectation of the system is null both before and immediately after the quench, hence the change in the system internal energy $\Delta U$ is null. Since there is no heat flow to the thermal environment either, $Q=0$ (because there is no time for that to happen), the work $W= \Delta U - Q$ is also null. Regarding the second quench, which is happening at the end of the protocol, its work cost is given by the change in energy across the quench. Since after the quench it is $H_0= 0$, the post quench energy expectation is trivially null, and the negative energy expectation $-U_f=-\langle H(\tau) \rangle_{\tau}$ over the final state gives the contribution of the quench to the total work. Since the final hamiltonian $H(\tau)$ contains only $\sigma^z$ terms, we can directly measure this contribution. Thus in the case of quantum cooperative erasure the total estimated work amounts to:
\begin{align}
W_\text{exp} = -U_f +W_z + W_{zz} \pm \delta W
\end{align}

Table \ref{tab:tab1} (column 4) reports the experimental values of $W_z, W_{zz}, U_f, \delta W, W_\text{exp}$ as well as the Landauer bound $W_L=N kT \ln 2$ for the cooperative erasure of $N=256$ bits of information. 

Dividing by the number of qubits, the measured work per qubit associated to cooperative quantum information erasure was
\begin{align}
w_\text{exp} &= 3,55 (0,73) \times 10^{-18} \text{erg/bit} \label{eq:our-w}
\end{align}
which is consistent (within the measurement error) with Landauer minimal cost reading
\begin{align}
w_L &= k T \ln 2 = 3,7 (0,9) \times 10^{-18} \text{erg/bit} \label{eq:wL}
\end{align}
The latter was calculated based on our estimation of the temperature at which our erasure protocol occurred, which is $T = 39\pm 9$ mK.

\section{Discussion and Conclusions}

Multiplying the measured time  $\mathcal T_{99,9\%}$ 
by the associated work per qubit $w_\text{exp}$, we obtain the record erasure action of 
\begin{align}
\mathcal A_{99,9\%}&= 3.19(0.27) \times 10^{-23} \text{erg}\cdot\text{s/bit} 
\label{eq:A_R}
\end{align}
which to the best of our knowledge is the smallest reported so far in the literature.

In this regard it is worth remarking that for any erasing device, the erasure action alone does not fully capture the overall performance of the device because it does not contain the information about the success rate $R$. As mentioned above, there is a trade-off between erasure action and success rate. This is a fact that was not acknowledged in the previous literature but  is necessary for a fair comparison of distinct information erasing devices. 
For future studies, it is important that erasure actions of distinct devices and methods be compared only if they refer to same success rate. This is why we use the label $99,9\%$ in Eq. (\ref{eq:A_R}). Alternatively a figure of merit that accounts for the success rate value should be employed. 
One possibility is the following figure of merit 
\begin{align}
\mathcal A^*= w \mathcal T (1-R)\, .
\end{align}
With this choice, given two devices with same erasure action $w \mathcal T$, the one with highest success rate $R$ would have a lower $A^*$. 
Clearly this is not the only possible choice, and further studies are in order to assess what would be the best choice.

This said, it is fair to state that the present work has experimentally demonstrated a new information erasure method of exceptional performance. The method erases $N$ qubits at once with nearly minimal Landauer energy cost, short time (order of microseconds), and impressively high success rate ($99,9\%$).  This is possible thanks to a shift in cooling paradigm whereby SSB and quantum phenomena work in synergy.

We point out that our erasure protocol was not optimised (for example we did not investigate how the qubit network responds when the protocol path is traversed faster or gets deformed) therefore we believe that there is room to further improve its performance. 

Furthermore, because it is based on an emergent phenomenon occurring in the large $N$ limit (namely the mechanism of SSB), the method is expected to perform the better the larger $N$, without any modification in its complexity, time duration and energy cost per erased bit. 
We note that this expectation is at odds with the expectation, based on quantum information theory, that larger systems and/or larger rate of erasure success require increasingly complex or increasingly long protocols \cite{Taranto21arXiv:2106.05151}. Our results indicates that many body emergent phenomena (specifically SSB) which were not accounted so far in the context of quantum information erasure or quantum algorithmic cooling may be the key to sidestep such theoretical limitations. Further sets of erasure experiments, investigating the large $N$ asymptotics, are in order to substantiate this expectation.

We stress that at variance with algorithmic cooling methods, the present one cools down all qubits participating in the process, not just a fraction thereof. Basic thermodynamics then imply that the thermal environment gets heated up as a consequence of the information erasure. In this regard, and at variance with algorithmic cooling, it is important to remark that the present method crucially relies on the fact that the qubit network is an open quantum system: were it isolated and evolving unitarily the method would never work. Quite paradoxically, environmental noise is here found to be key for qubit purification.

Most importantly, the present work also shows how NISQs, i.e., noisy intermediate-scale quantum devices \cite{Preskill18QUANTUM2} constitute a tremendous new tool for discovering new physics in the quantum regime. Not only the noisy nature of the annealer employed here was crucial for discovering the phenomenon of cooperative quantum information erasure, but also the very possibility of accessing that device remotely and experimenting with it, made the discovery quick and not expensive. Predicting the new phenomenon via numerical simulations on standard computers would have been much more time (and energy) consuming, and yet would lack of experimental validation.

Lastly, because it is already demonstrated on a NISQ device, the method can be readily employed to effectively initialise many qubits  in a quantum state of high purity and long duration, on next generation hybrid gate-based/quantum-annealing quantum computers, which marks a significant technological development.

\section*{Acknowledgements} 
Access to D-Wave services was provided by CINECA through the competitive ISCRA-C project ``q-land''. LB gratefully acknowledges The Blanceflor Foundation for financial support through the project ``Large Deviations approach to Landauer's Principle (LanDev)''. The authors thank Yasser Omar for comments on the manuscript.

\bibliographystyle{plainnat}
\bibliography{bibliography}

\begin{thebibliography}{28}
\providecommand{\natexlab}[1]{#1}
\providecommand{\url}[1]{\texttt{#1}}
\expandafter\ifx\csname urlstyle\endcsname\relax
  \providecommand{\doi}[1]{doi: #1}\else
  \providecommand{\doi}{doi: \begingroup \urlstyle{rm}\Url}\fi

\bibitem[dwd(Visited on 2022)]{dwdocs}
\url{https://docs.dwavesys.com/docs/latest/index.html}, Visited on 2022.

\bibitem[Auff\`eves(2022)]{Auffeves22PRXQ3}
Alexia Auff\`eves.
\newblock Quantum technologies need a quantum energy initiative.
\newblock \emph{PRX Quantum}, 3:\penalty0 020101, 2022.
\newblock \doi{10.1103/PRXQuantum.3.020101}.

\bibitem[Baugh et~al.(2005)Baugh, Moussa, Ryan, Nayak, and
  Laflamme]{Baugh05NAT438}
J.~Baugh, O.~Moussa, C.~A. Ryan, A.~Nayak, and R.~Laflamme.
\newblock Experimental implementation of heat-bath algorithmic cooling using
  solid-state nuclear magnetic resonance.
\newblock \emph{Nature}, 438\penalty0 (7067):\penalty0 470--473, 2005.
\newblock \doi{10.1038/nature04272}.
\newblock URL \url{https://doi.org/10.1038/nature04272}.

\bibitem[Benedetti et~al.(2016)Benedetti, Realpe-G{\'o}mez, Biswas, and
  Perdomo-Ortiz]{benedetti2016estimation}
Marcello Benedetti, John Realpe-G{\'o}mez, Rupak Biswas, and Alejandro
  Perdomo-Ortiz.
\newblock Estimation of effective temperatures in quantum annealers for
  sampling applications: A case study with possible applications in deep
  learning.
\newblock \emph{Physical Review A}, 94\penalty0 (2):\penalty0 022308, 2016.
\newblock \doi{10.1103/PhysRevA.94.022308}.
\newblock URL \url{https://link.aps.org/doi/10.1103/PhysRevA.94.022308}.

\bibitem[B{\'{e}}rut et~al.(2013)B{\'{e}}rut, Petrosyan, and
  Ciliberto]{Berut13EPL103}
A.~B{\'{e}}rut, A.~Petrosyan, and S.~Ciliberto.
\newblock Detailed jarzynski equality applied to a logically irreversible
  procedure.
\newblock 103:\penalty0 60002, 2013.
\newblock \doi{10.1209/0295-5075/103/60002}.

\bibitem[Berut et~al.(2012)Berut, Arakelyan, Petrosyan, Ciliberto,
  Dillenschneider, and Lutz]{Berut12Nature483}
Antoine Berut, Artak Arakelyan, Artyom Petrosyan, Sergio Ciliberto, Raoul
  Dillenschneider, and Eric Lutz.
\newblock Experimental verification of {L}andauer's principle linking
  information and thermodynamics.
\newblock \emph{Nature}, 483\penalty0 (7388):\penalty0 187--189, 2012.
\newblock \doi{https://doi.org/10.1038/nature10872}.

\bibitem[B{\'{e}}rut et~al.(2015)B{\'{e}}rut, Petrosyan, and
  Ciliberto]{Berut15JSTAT}
Antoine B{\'{e}}rut, Artyom Petrosyan, and Sergio Ciliberto.
\newblock Information and thermodynamics: experimental verification of
  {L}andauer{\textquotesingle}s erasure principle.
\newblock \emph{J. Stat. Mech.: Theory Exp.}, 2015:\penalty0 P06015, 2015.
\newblock \doi{10.1088/1742-5468/2015/06/p06015}.

\bibitem[Buffoni and Campisi(2022)]{Buffoni22JSP186}
Lorenzo Buffoni and Michele Campisi.
\newblock Spontaneous fluctuation-symmetry breaking and the {L}andauer
  principle.
\newblock \emph{J. Stat. Phys.}, 186\penalty0 (2):\penalty0 31, 2022.
\newblock \doi{10.1007/s10955-022-02877-8}.
\newblock URL \url{https://doi.org/10.1007/s10955-022-02877-8}.

\bibitem[Buffoni et~al.(2022)Buffoni, Gherardini, Cruzeiro, and
  Omar]{buffoni2022third}
Lorenzo Buffoni, Stefano Gherardini, Emmanuel~Zambrini Cruzeiro, and Yasser
  Omar.
\newblock Third law of thermodynamics and the scaling of quantum computers.
\newblock \emph{Phys. Rev. Lett.}, \penalty0 (129):\penalty0 150602, 2022.
\newblock \doi{10.1103/PhysRevLett.129.150602}.
\newblock URL \url{https://link.aps.org/doi/10.1103/PhysRevLett.129.150602}.

\bibitem[Cirac et~al.(1999)Cirac, Ekert, and Macchiavello]{Cirac99PRL82}
J.~I. Cirac, A.~K. Ekert, and C.~Macchiavello.
\newblock Optimal purification of single qubits.
\newblock \emph{Phys. Rev. Lett.}, 82:\penalty0 4344--4347, 1999.
\newblock \doi{10.1103/PhysRevLett.82.4344}.

\bibitem[Esposito et~al.(2010)Esposito, Lindenberg, and den
  Broeck]{Esposito10NJP12}
Massimiliano Esposito, Katja Lindenberg, and Christian~Van den Broeck.
\newblock Entropy production as correlation between system and reservoir.
\newblock \emph{New J. Phys.}, 12:\penalty0 013013, 2010.
\newblock \doi{10.1088/1367-2630/12/1/013013}.

\bibitem[Fellous-Asiani et~al.(2021)Fellous-Asiani, Chai, Whitney,
  Auff{\`e}ves, and Ng]{fellous2021limitations}
Marco Fellous-Asiani, Jing~Hao Chai, Robert~S Whitney, Alexia Auff{\`e}ves, and
  Hui~Khoon Ng.
\newblock Limitations in quantum computing from resource constraints.
\newblock \emph{PRX Quantum}, 2\penalty0 (4):\penalty0 040335, 2021.
\newblock \doi{10.1103/PRXQuantum.2.040335}.
\newblock URL \url{https://link.aps.org/doi/10.1103/PRXQuantum.2.040335}.

\bibitem[Fernandez et~al.(2004)Fernandez, Lloyd, Mor, and
  Roychowdhury]{Fernandez04IJQI02}
Jos\'e Fernandez, Seth Lloyd, Tal Mor, and Vwani Roychowdhury.
\newblock Algorithmic cooling of spins: a practicable method for increasing
  polarization.
\newblock \emph{International Journal of Quantum Information}, 02:\penalty0
  461--477, 2004.
\newblock URL \url{https://doi.org/10.1142/S0219749904000419}.

\bibitem[Gaudenzi et~al.(2018)Gaudenzi, Burzur{\'\i}, Maegawa, van~der Zant,
  and Luis]{Gaudenzi18NATPHYS14}
R.~Gaudenzi, E.~Burzur{\'\i}, S.~Maegawa, H.~S.~J. van~der Zant, and F.~Luis.
\newblock Quantum {L}andauer erasure with a molecular nanomagnet.
\newblock \emph{Nat. Phys.}, 14\penalty0 (6):\penalty0 565--568, 2018.
\newblock \doi{10.1038/s41567-018-0070-7}.

\bibitem[Gavrilov and Bechhoefer(2016)]{Gavrilov16PRL117}
Mom{\v{c}}ilo Gavrilov and John Bechhoefer.
\newblock Erasure without work in an asymmetric double-well potential.
\newblock \emph{Phys. Rev. Lett.}, 117:\penalty0 200601, 2016.
\newblock \doi{10.1103/PhysRevLett.117.200601}.

\bibitem[{Gavrilov} et~al.(2017){Gavrilov}, {Ch{\'e}trite}, and
  {Bechhoefer}]{Gavrilov17PNAS114}
Mom{\v{c}}ilo {Gavrilov}, Rapha{\"e}l {Ch{\'e}trite}, and John {Bechhoefer}.
\newblock {Direct measurement of weakly nonequilibrium system entropy is
  consistent with Gibbs-Shannon form}.
\newblock \emph{Proceedings of the National Academy of Science}, 114\penalty0
  (42):\penalty0 11097--11102, 2017.
\newblock \doi{10.1073/pnas.1708689114}.

\bibitem[Goldenfeld(1992)]{GoldenfeldBook}
N.~Goldenfeld.
\newblock \emph{Lectures On Phase Transitions And The Renormalization Group}.
\newblock CRC Press, Boca Raton, 1st edition, 1992.
\newblock \doi{https://doi.org/10.1201/9780429493492}.

\bibitem[Landauer(1961)]{Landauer61IBMRD5}
Rolf Landauer.
\newblock Irreversibility and heat generation in the computing process.
\newblock \emph{IBM J. Res. Dev.}, 5\penalty0 (3):\penalty0 183--191, 1961.
\newblock \doi{10.1147/rd.53.0183}.

\bibitem[Leonard J.~Schulman(1999)]{Schulman99STOC99}
Umesh V.~Vazirani Leonard J.~Schulman.
\newblock Molecular scale heat engines and scalable quantum computation.
\newblock In \emph{STOC '99 Proceedings of the thirty-first annual ACM
  symposium on theory of computing}, pages 322--329, 1999.
\newblock \doi{10.1145/301250.301332}.

\bibitem[Likharev(1977)]{Likharev77IEEETM13}
K.~Likharev.
\newblock Dynamics of some single flux quantum devices: I. parametric quantron.
\newblock \emph{IEEE Transactions on Magnetics}, 13\penalty0 (1):\penalty0
  242--244, 1977.
\newblock \doi{10.1109/TMAG.1977.1059351}.

\bibitem[Parrondo(2001)]{Parrondo01CHAOS11}
Juan M.~R. Parrondo.
\newblock The {S}zilard engine revisited: Entropy, macroscopic randomness, and
  symmetry breaking phase transitions.
\newblock \emph{Chaos}, 11\penalty0 (3):\penalty0 725--733, 2001.
\newblock \doi{10.1063/1.1388006}.

\bibitem[Piechocinska(2000)]{Piechocinska00PRA61}
Barbara Piechocinska.
\newblock Information erasure.
\newblock \emph{Phys. Rev. A}, 61:\penalty0 062314, May 2000.
\newblock \doi{10.1103/PhysRevA.61.062314}.
\newblock URL \url{https://link.aps.org/doi/10.1103/PhysRevA.61.062314}.

\bibitem[Preskill(2018)]{Preskill18QUANTUM2}
John Preskill.
\newblock Quantum {C}omputing in the {NISQ} era and beyond.
\newblock \emph{{Quantum}}, 2:\penalty0 79, 2018.
\newblock \doi{10.22331/q-2018-08-06-79}.

\bibitem[Proesmans et~al.(2020)Proesmans, Ehrich, and
  Bechhoefer]{Proesmans20PRL125}
Karel Proesmans, Jannik Ehrich, and John Bechhoefer.
\newblock Finite-time landauer principle.
\newblock \emph{Phys. Rev. Lett.}, 125:\penalty0 100602, 2020.
\newblock \doi{10.1103/PhysRevLett.125.100602}.

\bibitem[Saira et~al.(2020)Saira, Matheny, Katti, Fon, Wimsatt, Crutchfield,
  Han, and Roukes]{Saira20PRR2}
Olli-Pentti Saira, Matthew~H. Matheny, Raj Katti, Warren Fon, Gregory Wimsatt,
  James~P. Crutchfield, Siyuan Han, and Michael~L. Roukes.
\newblock Nonequilibrium thermodynamics of erasure with superconducting flux
  logic.
\newblock \emph{Phys. Rev. Research}, 2:\penalty0 013249, 2020.
\newblock \doi{10.1103/PhysRevResearch.2.013249}.

\bibitem[Schulman et~al.(2005)Schulman, Mor, and Weinstein]{Schulman05PRL94}
Leonard~J. Schulman, Tal Mor, and Yossi Weinstein.
\newblock Physical limits of heat-bath algorithmic cooling.
\newblock \emph{Phys. Rev. Lett.}, 94:\penalty0 120501, 2005.
\newblock \doi{10.1103/PhysRevLett.94.120501}.

\bibitem[Solfanelli et~al.(2022)Solfanelli, Santini, and
  Campisi]{Solfanelli22AVSQS4}
Andrea Solfanelli, Alessandro Santini, and Michele Campisi.
\newblock Quantum thermodynamic methods to purify a qubit on a quantum
  processing unit.
\newblock \emph{AVS Quantum Science}, 4\penalty0 (2):\penalty0 026802, 2022.
\newblock \doi{10.1116/5.0091121}.

\bibitem[{Taranto} et~al.(2021){Taranto}, {Bakhshinezhad}, {Bluhm}, {Silva},
  {Friis}, {Lock}, {Vitagliano}, {Binder}, {Debarba}, {Schwarzhans}, {Clivaz},
  and {Huber}]{Taranto21arXiv:2106.05151}
Philip {Taranto}, Faraj {Bakhshinezhad}, Andreas {Bluhm}, Ralph {Silva},
  Nicolai {Friis}, Maximilian P.~E. {Lock}, Giuseppe {Vitagliano}, Felix~C.
  {Binder}, Tiago {Debarba}, Emanuel {Schwarzhans}, Fabien {Clivaz}, and Marcus
  {Huber}.
\newblock {Landauer vs. Nernst: What is the True Cost of Cooling a Quantum
  System?}
\newblock \emph{arXiv:2106.05151}, 2021.
\newblock URL \url{https://doi.org/10.48550/arXiv.2106.05151}.

\end{thebibliography}

\bigskip

\end{document}